\documentclass[12pt,preprint]{aastex}
\newcommand{\E}{\times 10^}
\begin{document}
\title{Luminosity Function of Faint Globular Clusters in M87}
\author{Christopher Z. Waters and Stephen E. Zepf}
\affil{Department of Physics and Astronomy, Michigan State University, 3246 
  Biomedical Physical Sciences, East Lansing, MI 48824}
\email{watersc1@pa.msu.edu,zepf@pa.msu.edu}
\author{Tod R. Lauer}
\affil{National Optical Astronomy Observatory, P. O. Box 26732, 
  Tucson, AZ 85726}
\email{lauer@noao.edu}
\author{Edward A. Baltz}
\affil{KIPAC, Stanford University, P. O. Box 20450, MS 29, Stanford, CA 94309}
\email{eabaltz@slac.stanford.edu}
\author{Joseph Silk}
\affil{University of Oxford, Astrophysics, Keble Road, Oxford OX1 3RH, 
  United Kingdom}
\email{silk@astro.ox.ac.uk}

\begin{abstract}
  We present the luminosity function to very faint magnitudes for the
  globular clusters in M87, based on a 30 orbit \textit{Hubble Space
    Telescope (HST)} WFPC2 imaging program. The very deep images and
  corresponding improved false source rejection allow us to probe the
  mass function further beyond the turnover than has been done before.
  We compare our luminosity function to those that have been observed
  in the past, and confirm the similarity of the turnover luminosity
  between M87 and the Milky Way.  We also find with high statistical
  significance that the M87 luminosity function is broader than that
  of the Milky Way.  We discuss how determining the mass function of the
  cluster system to low masses can constrain theoretical models of the
  dynamical evolution of globular cluster systems.  Our mass function is
  consistent with the dependence of mass loss on the initial cluster
  mass given by classical evaporation, and somewhat inconsistent with 
  newer proposals that have a shallower mass dependence.  In addition,
  the rate of mass loss is consistent with standard evaporation
  models, and not with the much higher rates proposed by some recent
  studies of very young cluster systems.  We also find that the
  mass-size relation has very little slope, indicating that there is
  almost no increase in the size of a cluster with increasing mass.
\end{abstract}

\keywords{galaxies: star clusters -- galaxies: individual (M87) -- 
  globular clusters: general}

\section{Introduction}

The dynamical evolution of globular clusters has long been the subject
of many theoretical studies, including a number of recent papers
utilizing state of the art simulations \citep[e.g.][ and references
therein]{BM}.  These studies find that mass loss due to two body
relaxation and the resulting evaporation is the primary mechanism for
the destruction of globular clusters.  Only those clusters with
disruption timescales similar to or less than a Hubble time will be
significantly changed by this evolution over their lifetimes.  As
evaporation causes clusters to lose mass at a fairly constant rate,
the disruption timescale is directly related to the initial mass of
the cluster.  Therefore, the clusters with the lowest initial masses
should constrain the form of this evolution most strongly.  These low
mass clusters have lower luminosities, which makes them difficult to
observe in other galaxies.  For the Milky Way, where finding lower
luminosity clusters is not as difficult, the small total number of
clusters prevents any results from being statistically strong.  What
is needed to provide constraints on the models of the dynamical
evolution of globular clusters is a system which has both large
numbers, and observations that reach low masses with a high degree of
accuracy and reliability.

In this paper, we present results from new, very deep HST images of
M87 that achieves these goals of depth and large numbers of globular
clusters.  In \S 2, we present the data and explain the procedures
used to measure the globular clusters in M87.  We present our
luminosity function in \S 3, and compare it to other deep luminosity
functions that have been previously published.  We show our mass
function, and show the constraints it gives for theoretical
models of globular cluster evolution in \S 4.  Finally, we present our
conclusions in \S 5.

\section{Observations and Reductions}

\subsection{Observations and Data}
The center of M87 was imaged with the WFPC2 as part of a 30 orbit
program \citep{Baltz} in May and June of 2001.  During each orbit, four
exposures totalling 1040s were taken in the F814W filter, with a
single matching 400s exposure in the F606W filter.  The four F814W
images were dithered by half pixel steps to allow for the images to be
interlaced directly into a 2x2 grid.  Average images were created from
the final full data set by stacking the images, after they had been
sinc-interpolated to a common origin.  The images from day 12 were
excluded from this step as they had more jitter than the
other images.  This combination creates images that have total
exposure times of $30160$s for F814W, and $11600$s for F606W.

Cluster detection was performed separately on the images for each
filter using Source Extractor \citep{sextractor}.  As we are imaging
the center of the galaxy, the galaxy profile is steep.  Because the
main source of noise in the image is from the variation in the number
of photons detected from the galaxy light, this steep profile creates
a rapid change in the noise level across the images.  To prevent the
change in noise from altering how clusters are detected at different
distances from the galaxy, we created an image to model the background
level of the galaxy itself.  This was done by median filtering the
image with a box larger than the size of any of the globular cluster
features.  As the radii of the largest clusters can be up to 20
pixels, we use a filter box of 50 x 50 pixels.  This background image
was used to weight the detection threshold, which helps minimize
radial variations in our object detection.

\subsection{Object Selection and Completeness}
Since the faint end of the observed GCLF is sensitive to the
contamination of objects that are not globular clusters, we want to
ensure that we remove as many of these objects as possible.  The
majority of the very faintest objects detected are likely to be simply
peaks in the variation of the background.  As the background
subtracted image histograms are symmetric about zero, we expect that
the magnitude and frequency of these variations are the same between
peaks and valleys.  Therefore, by inverting the sign of the data
images (simply multiplying by $-1$), and searching on these images, we
can determine how many candidate clusters are caused by the variations
in the background level.  Based on these tests, we chose detection
thresholds of $3\sigma$ with a minimum object size of 2 pixels.
Although it is tempting to push this lower to increase the completion
at fainter magnitudes, below this limit the increase in the number of
objects detected on the data image is closely matched by an increase
in the number of objects detected on the inverted image.  This
suggests that we are not reliably adding new real clusters to our
sample.

In addition to these false clusters, we also expect to have a
contribution from faint galaxies behind M87.  This contribution can be
estimated by measuring how many objects we find with our settings on
very deep HST images like those of the Hubble Deep Field \citep{HDF}.
The detection efficiency for faint extended objects is very sensitive
to the signal to noise ratio of the image.  This required us to
degrade the quality of the HDF image to match the average variance of
our images.

To help remove as many contaminating objects as possible, we demand
that two requirements be met before an object was determined to be a
valid globular cluster.  The first requirement is that all globular
cluster candidates must be detected in both filters.  This removes a
significant portion of the false clusters, as well as a few of the HDF
objects.  By further requiring that all candidates have a final $V -
I$ color between 0.5 and 1.7, we exclude almost all of the remaining
background sources and false clusters.  As shown in Table
\ref{tab:detection}, each of these cuts removes approximately the same
number of clusters from both the observed and false sets which
suggests that these cuts preserve the real clusters, and only remove
objects that are not globular clusters.

With the detection parameters set, we next used model clusters of
known flux to determine the completeness of our sample.  All of the
model clusters were generated from King model profiles with a fixed
concentration of 1.26 at eight tidal radius steps evenly spaced
between 2 and 16 pixels.  These templates were then scaled to form a
grid in magnitude and size, with F814W and F606W apparent magnitudes
spaced every 0.5 magnitude between 20 and 28 magnitude.  This range of
radii and magnitudes was chosen to cover the properties of the real
clusters in our data sample.

For each combination of magnitude and tidal radius, 150 copies of the
template were randomly placed on the background subtracted images.
Source Extractor was then used to find these objects with the same
detection parameters as were used on the data image.  By measuring the
rate at which the templates are recovered, we can determine how
complete our data sample is.  Because the change in the completeness
did not vary much for template clusters of different sizes, the final
completeness is found by simply averaging the results for all radii
together.

As the detection process is sensitive to the background uncertainty,
the completeness limit becomes fainter as the distance from the center
of the galaxy increases.  To take advantage of this increased
sensitivity, we divided our sample into two radial bins with equal
numbers of clusters in each.  This gives bins that have rages of
projected galactocentric distances from $1.1$kpc to $4.4$kpc, and from
$4.4$kpc to $9.0$kpc.  The completeness for each bin was then
calculated based only on the templates that fell within that bin.

Our requirement that clusters be detected in both the F814W and F606W
filters also influences the final completeness curves, which are 
simply the product of the two completenesses for each filter.  For
clusters that are bright enough in both filters to be reliably
detected, this has no effect.  Because of the depth of our data, this
limit extends to $V \sim 25$, much fainter than the turnover of the
globular cluster luminosity function.  For clusters fainter than this,
the color dependence begins to exclude the bluest clusters.  The
resulting 50\% completeness limit for our samples in the two radius
bins is $V_{inner} = 25.07 + 0.62 (V - I)$ and $V_{outer} = 25.27 +
0.63(V - I)$.

\begin{figure}
  \figurenum{1}
  \plotone{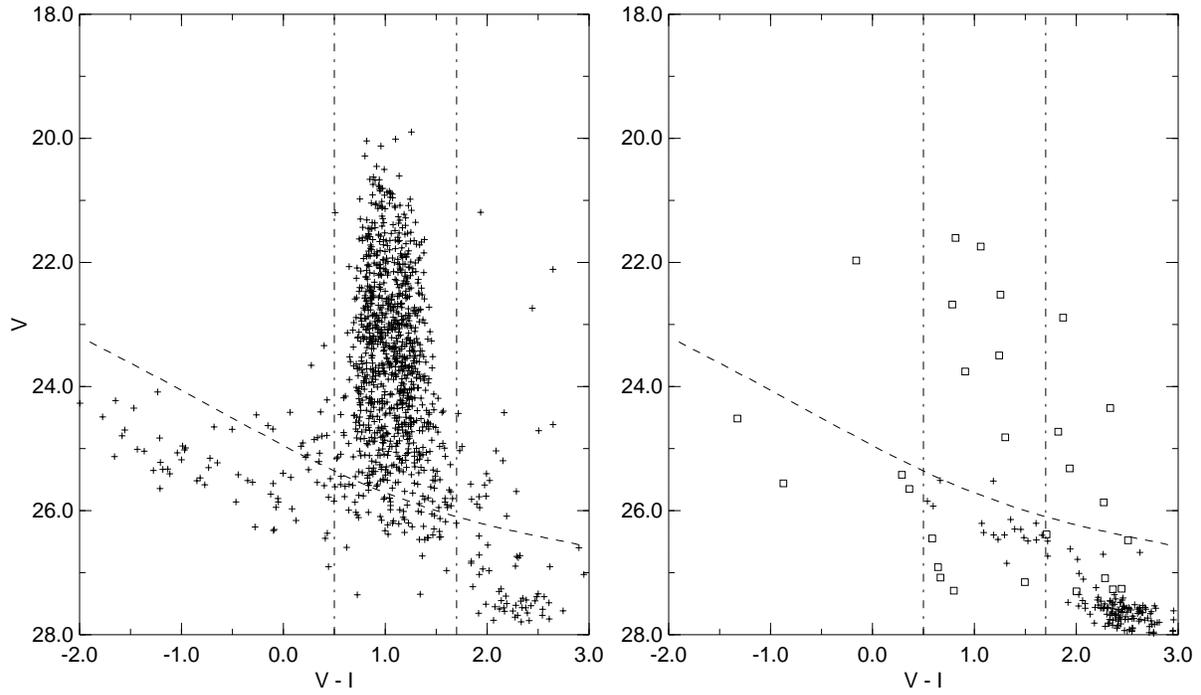}
  \caption{Observed color magnitude diagram.  The left panel shows the
    objects that were detected on our data images, and the right shows
    the results from our contamination tests, with the detections on
    the inverted images shown as crosses, and the detections from the
    HDF shown as boxes.  The dashed line is the median $50\%$
    completeness limit, and the dot dashed lines are the color limits
    chosen to select out globular clusters from the candidates.  
    \label{fig:color_mag}}
\end{figure}

Figure \ref{fig:color_mag} shows the final observed color magnitude
diagram, based on the photometric calibration discussed in section 2.3.
The left panel shows all objects detected in our data image, with the
right panel showing the detections from the HDF background field, as
well as from the image background fluctuations.  It is clear from this
plot that with the cuts that we have applied, nearly all of the 
contaminating objects are removed from the sample, leaving a sample
that is composed overwhelmingly by globular clusters.

\subsection{Aperture Correction and Photometric Calibration}

In addition to allowing us to measure the completeness of our detection
process, our templates also provide a direct way to measure the
aperture corrections for our clusters.  The globular clusters on our
image are extended objects, and as such, fixed apertures may not
measure all the light equally well for all clusters.  We can estimate
what the aperture correction should be for our clusters by comparing
the measured magnitudes for the templates to the known magnitude with
which they were created.

All photometry was done using two fixed apertures of 2 and 4 pixels.
We were then able to construct a parameter to estimate how extended
the clusters are by taking the difference between these two
magnitudes.  By taking the median values for this parameter, the outer
aperture magnitude, and the required aperture correction for each
combination of template size and input magnitude, we were able to
create a mesh of values that can be used to calibrate the real
clusters.  For each cluster, the aperture correction was calculated by
finding the value at the mesh point that most closely matched the
observed outer aperture magnitude and magnitude difference.  This
method works well for most of the observed clusters, although at faint
magnitudes, the scatter in our measured parameters increases, and the
decrease in the number of templates recovered allows this increased
scatter to propagate to the mesh point values.  By passing the
template clusters through this procedure as well as the real clusters,
we were able to determine that the aperture correction introduces an
uncertainty of 0.1 magnitudes into the final magnitudes, similar to
what is found in other HST studies of extragalactic globular clusters
\citep[e.g.][]{Kundu}.  As both filters are affected equally by this
uncertainty, the calculated colors are not changed.

We used the zeropoints and color corrections listed in
\citet{Holtzman} to convert the aperture corrected instrumental
magnitudes into standard V and I magnitudes.  The extinction for M87
from \citet{Schlegel} was also applied to our measurements.  Finally,
the standard 0.1 magnitude aperture correction was applied to the
final magnitude to correct for the light that is lost to scattering by
the WFPC2 optics.  To confirm that our magnitudes match those
previously published for the M87 globular clusters, we compared our
final magnitudes to those presented by \citet{Kundu} for the 346
clusters that are common to both samples. The average differences
between our magnitudes for the clusters are $\langle V -
V_{Kundu}\rangle = -0.042$ and $\langle I - I_{Kundu}\rangle =
-0.021$, well within our $0.1$ magnitude uncertainty due to the
aperture correction.

\begin{figure}
  \figurenum{2}
  \plotone{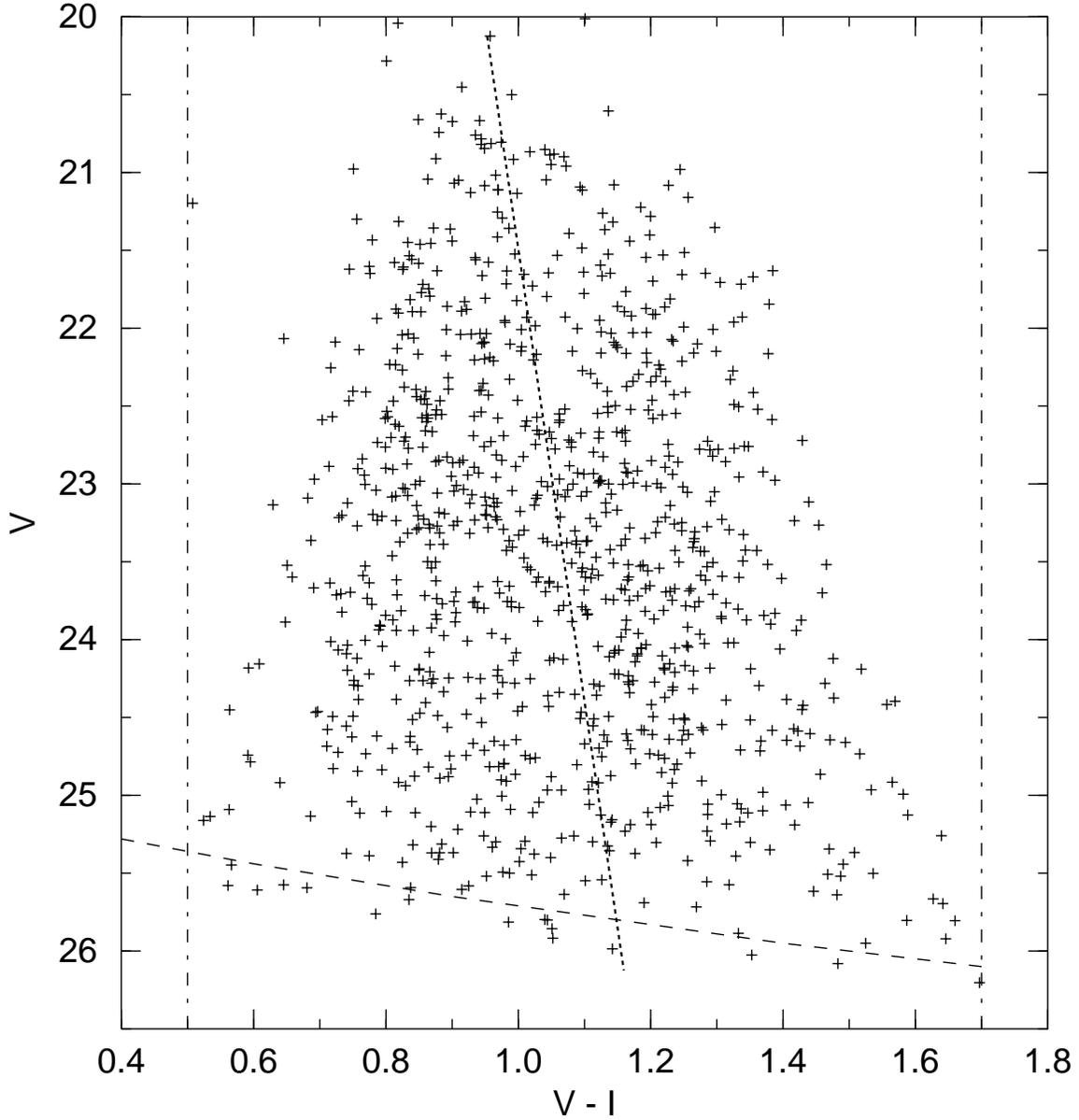}
  \caption{Close up of the range of color and magnitude considered for
    globular clusters.  The color cuts and median $50\%$ completeness
    limits are the same as in figure \ref{fig:color_mag}.  The weak trend
    between the median brightness of the cluster with the cluster
    color is plotted as a dotted line.    \label{fig:color_mag2}}
\end{figure}

The color magnitude diagram for the objects we consider globular
clusters is presented in figure \ref{fig:color_mag2}.  This sample
contains only the points that fall within our color cuts, as well as
to detections that are above the 50\% completeness limit.  We have
also overplotted a line showing the weak trend between the median
brightness of cluster as a function of color, $V \propto 29 (V - I)$.
This difference follows naturally from stellar population models,
which predict that the lower metallicity blue clusters will have a
smaller $M / L_V$ than the higher metallicity red ones \citep[see
discussion in][and references therein]{Kundu}.  This also creates a
spread in the $M / L_V$ ratios for the clusters, which tends to
broaden the luminosity function.  

\subsection{Size Measurements}
\begin{figure}
  \figurenum{3}
  \plotone{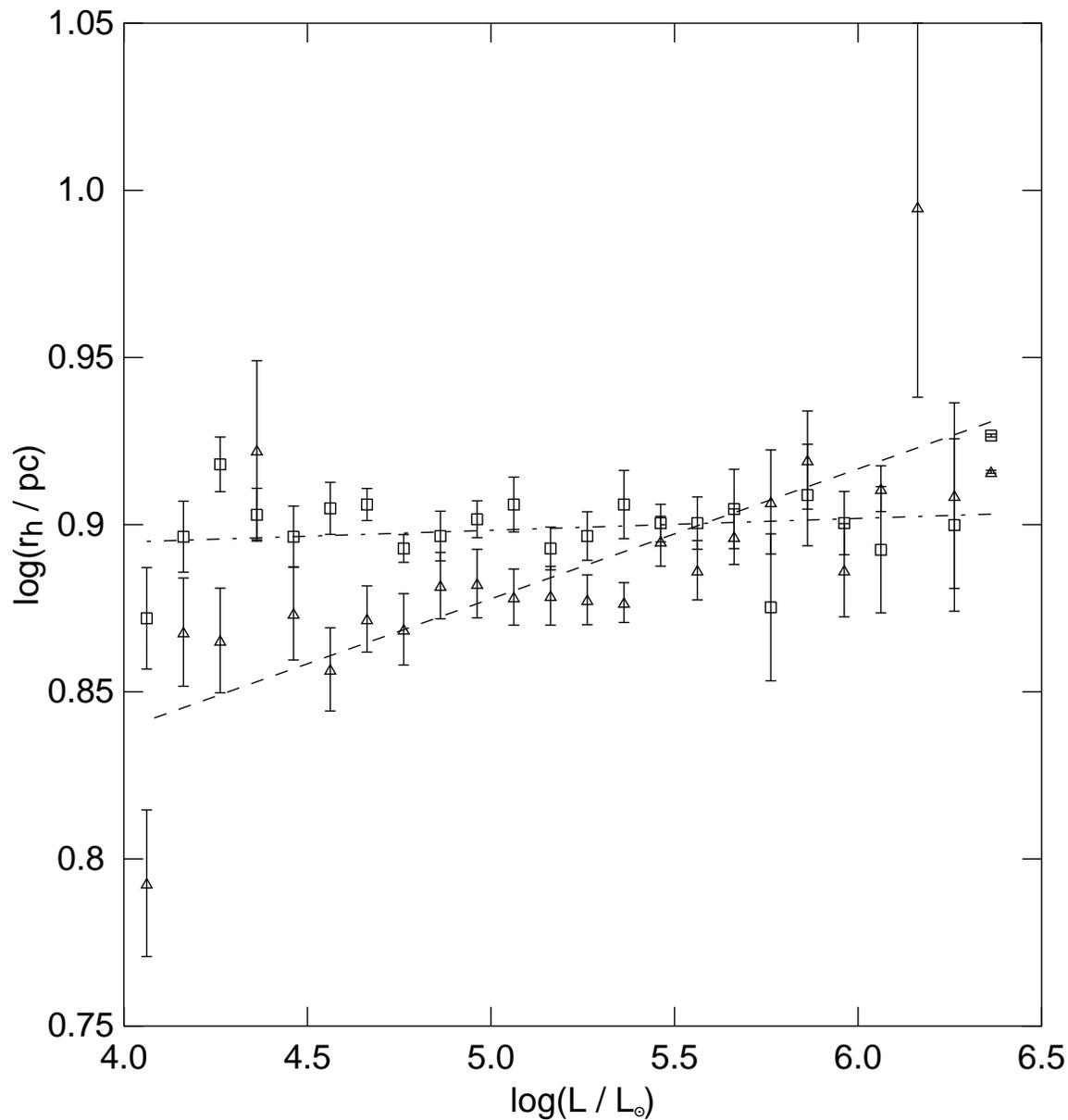}
  \caption{Relation between the observed half light radius of the
    globular clusters and their luminosity.  The triangles illustrate
    the median sizes for each luminosity bin, and the squares are the
    sizes determined from a random sample that has been clipped based
    on the completeness function observed.  The best fitting relations
    are plotted for both the real half-light radius data (dashed line,
    $r_{half} \propto L^{0.04}$) and the random sample of data
    (dot-dashed line, $r_{half} \propto L^{0.01}$).
    \label{fig:size_mag}}
\end{figure}

The half-light radii for each cluster was measured using Source
Extractor.  As shown in Figure \ref{fig:size_mag}, the measured range
of half-light radii for the real clusters appears to be generally
constant over all magnitudes.  This suggests that any biases in our
measurements of the cluster sizes must be fairly small.  To confirm
this, we generated a random sample that is evenly populated over the
observed range of $V-I$ color, $V$ magnitude and $r_{half}$ sizes.  At
each point, we calculated the completeness as a function of these
three parameters.  The observed half-light radius was included by
keeping the completeness results separated by the input template
sizes.  We then assigned that size template the median half-light
radius that was found during the template recovery.

Using this sample, we can investigate how strongly we expect the real
cluster sample to be biased.  The photometric errors for the faintest
clusters should tend to decrease the measured half-light radius.  In
addition, the low surface brightness of large faint clusters may also
cause such clusters to not be detected, which will also decrease the
average size observed for faint clusters.  We selected only those
clusters that had calculated completeness greater than 50\%.  The
median half-light radius was then found for bins spaced equally in
$\log(L)$ for both the random sample and the sample of real clusters,
with errors determined via bootstrapping.  Linear fits were then
performed on the values of $\log(L)$ and $\log(r_{half})$.  These best
fitting lines, as well as the binned data are shown in Figure
\ref{fig:size_mag}.  We find that the random sample is best fit by a
relation $r_{half} \propto L^{0.01}$. This indicates that we are not
significantly biased against large, faint clusters, and that the
change in size due to errors must also be small.

For the real clusters, the trend is not much stronger, with $r_{half}
\propto L^{0.04}$.  The increased scatter compared to the random
sample is likely due to the real distribution of cluster
concentrations.  This introduces a distribution of half-light radii
that is larger than that in our random sample, as our use of only a
single concentration value in our templates ignores any such
variation.  The weak observed trend shows clearly that the median
half-light radius of globular clusters changes very little over the
magnitude range that we observe.  This result is similar to what has
been observed for the Milky Way system \citep[e.g.][]{AZ98}, as well
as for young globular cluster systems in other galaxies
\citep{Z1999,Larsen}.  The similarity between both the young and old
cluster systems suggests that this trend arises during cluster
formation, and as such, is an important constraint on formation models
for globular clusters \citep[see][and references therein]{AZ01}.

\section{Luminosity Function}

\begin{figure}
  \figurenum{4}
  \plotone{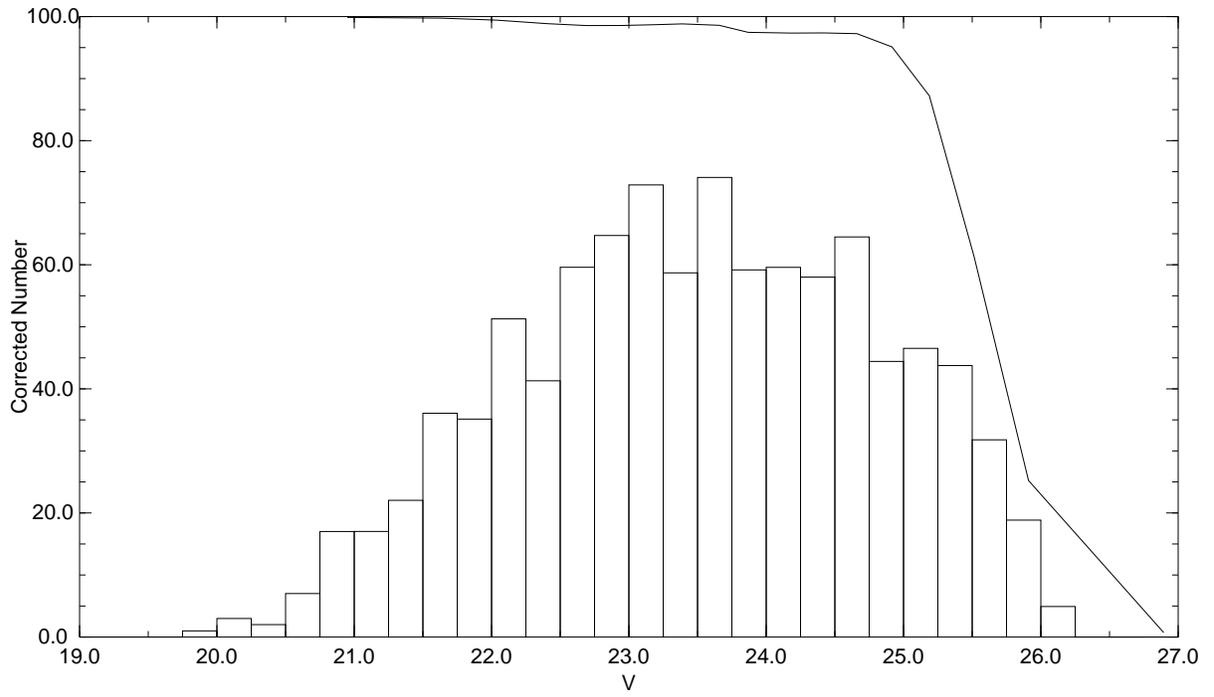}
  \caption{Completeness corrected luminosity function for the globular
    clusters.  The dashed line represents the median completeness as a
    function of magnitude. \label{fig:GCLF}}
\end{figure}

The luminosity function histogram (figure \ref{fig:GCLF}) was
created by binning each cluster by a weight factor, defined to be $W =
(\textup{completeness}(V,V-I))^{-1}$.  The error for each bin is then
$\langle W \rangle \sqrt{N}$, where $N$ is the number of clusters
actually detected in the bin, and $\langle W \rangle$ is the average
of the weight factors for the clusters in the bin.  The effect of the
contamination from false detections of background noise, as well as
from background galaxies, are weighted in the same way, and subtracted
from the bin that they would fall in if they were real clusters.  This
corrects our luminosity function both for any real clusters that we
may have missed, as well as for any objects detected that are not real
globular clusters.

\begin{figure}
  \figurenum{5}
  \plotone{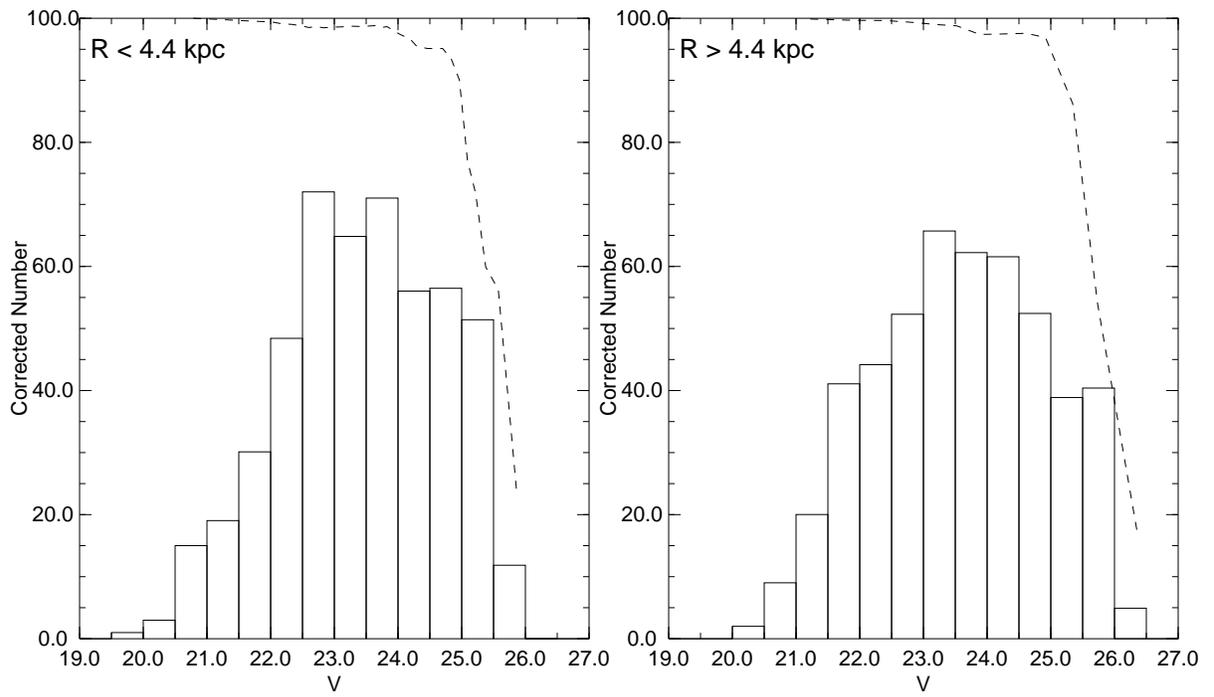}
  \caption{Completeness corrected luminosity functions for our inner
    and outer radius bins, with the boundary between the bins at 4.4
    kpc.  The median completeness as a function of magnitude for each
    sample is plotted as a solid line.  \label{fig:GCLF_R}}
\end{figure}

For each of our two radial bins, we created separate luminosity
functions in the same way as for the entire sample.  These two
luminosity functions are shown along with the median completeness for
each sample in figure \ref{fig:GCLF_R}.  By comparing these luminosity
functions for the two bins, we can check that any radial trend,
including from our completeness correction, does not influence the
final luminosity function.  This comparison was done on the unbinned
distributions by using an F-test to check that the variances are the
same between the two bins, and a t-test to check that the turnover in
the luminosity function is in the same location.  For both tests, we
use the hypothesis that the distributions are the same, and find that
we can accept the hypothesis for both tests (F-test p-value = 0.24;
t-test p-value = 0.94), indicating no significant difference between
the luminosity functions at our two radii.

Since the two radial bins are similar, we can work with the final
luminosity function at all radii, and compare this final luminosity
function to those that have been presented before for globular cluster
systems.  The two most important comparisons are to previous results
for the M87 globular clusters and to the Galactic globular cluster
system.  For M87, the most complete previous study of the globular
cluster luminosity function is by \citet{Kundu}.  They fit a Gaussian
to their binned luminosity function, finding that $\langle V \rangle =
23.67$ with $\sigma = 1.39$.  Repeating this for our luminosity
function yields similar results, with $\langle V \rangle = 23.60$ and
$\sigma = 1.42$.  We can also test how well the unbinned distributions
match, again using an F-test and a t-test to check the agreement of
the variances and turnover location.  The F-test allows us
to accept our hypothesis (p-value = 0.26), indicating that the
variances of the two distributions are consistent with each other.
The t-test also allows allows us to accept the hypothesis (p-value =
0.94), suggesting that the location of the turnover is the same for
both as well.  The combination of these two results leads to the
conclusion that our luminosity function is consistent with that
previously published for M87.

\begin{figure}
  \figurenum{6}
  \plotone{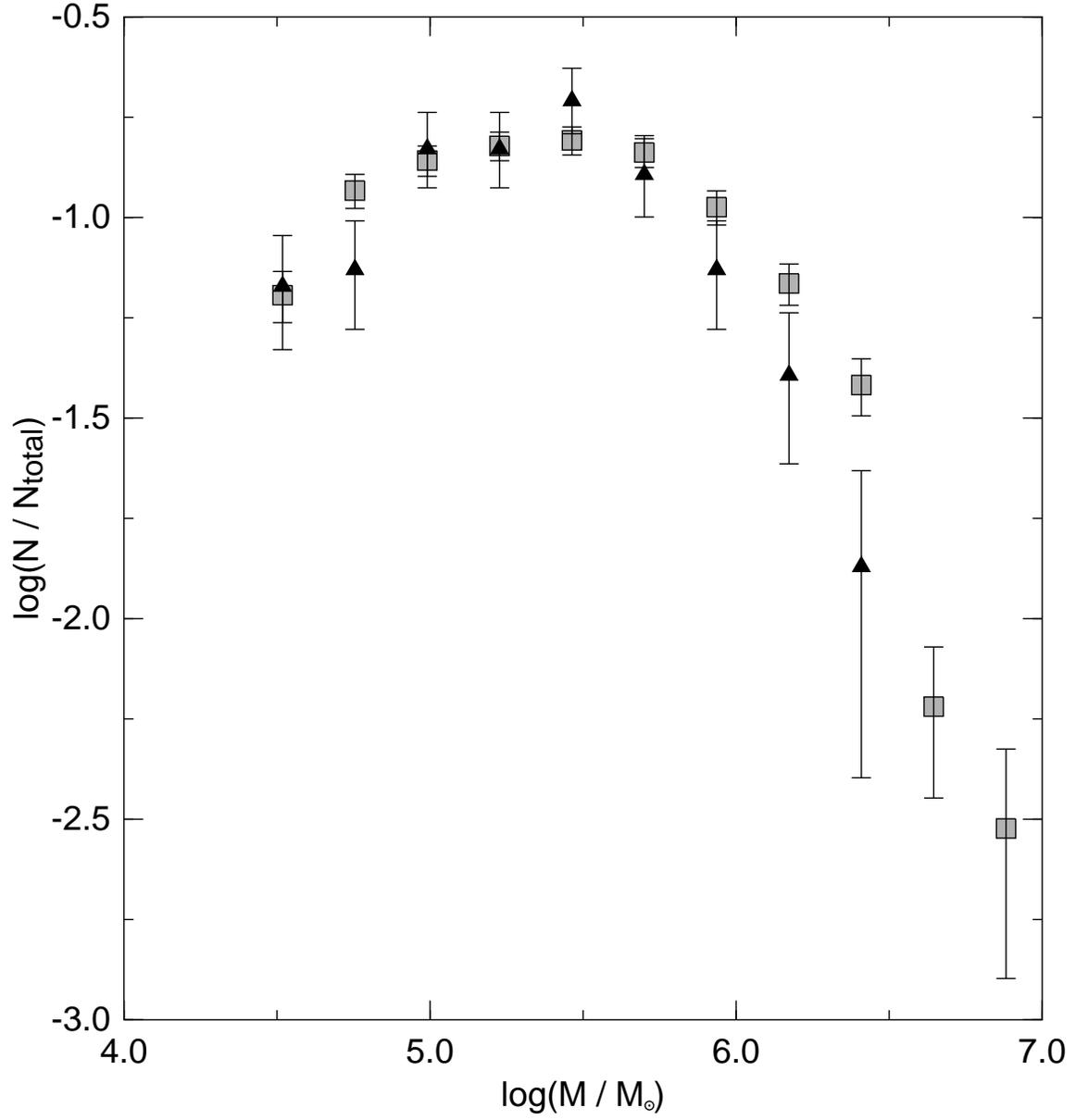}
  \caption{Comparison of our globular cluster mass function for M87
    (squares) to the mass function of the Milky Way (triangles), based
    on the globular cluster catalog of \citet{Harris}.
    \label{fig:mw_comp}}
\end{figure}

We can repeat this procedure for the Milky Way clusters, and see if
this luminosity function also matches.  We used the catalog of
galactic globular clusters of \citet{Harris}, and adopted a distance
to M87 of 16 Mpc \citep{macri}.  We find that a Gaussian fit to the
binned Milky Way luminosity function yields $\langle V \rangle =
23.70$ and $\sigma = 1.18$.  Running the same tests as above with the
same hypotheses, we can conclude that the turnovers are the same
(t-test p-value = 0.99), but that the variances are significantly
different (F-test p-value = $8\E{-3}$).  This can be seen in figure
\ref{fig:mw_comp}, which illustrates that the Milky Way distribution
is significantly narrower.  

This difference in the dispersions may be related to the differences
in the cluster color distributions between the galaxies.  M87 has a
strongly bimodal color distribution, with roughly equal numbers of red
and blue clusters.  These two populations serve to spread out the
total luminosity function, as the blue population tends to be brighter
than the red ($\langle V_{blue} \rangle = 23.27$, $\langle V_{red}
\rangle = 23.62$).  This separation serves to increase the total
luminosity function dispersion.  The shape of the luminosity function
remains a smooth Gaussian, as the dispersions of both of these
populations are larger than the separation in the means.  Because the
blue population dominates the Milky Way system, there is
correspondingly little effect from the spread of cluster colors on the
dispersion of the luminosity function.

\section{Mass function}

As our data probe well beyond the turnover of the GCLF, we can use it
to gauge how well current cluster evolution models track the
observations.  Theory predicts that dynamical evolution is the
dominant factor in the shape of the current luminosity function.  To
compare to theoretical models, we need to convert the luminosities
into masses.  We adopt a $M/L$ ratio of 3 based on work on the
structural parameters \citep{future} as our fiducial value.  Changes
in this value do not change the relative fit quality between the
observed mass function and the theoretical models that we compare to.
However, changes will alter the mass loss rates that we calculate from
our data, such that all of the rates that we calculate should be
scaled by a factor $(M / L) / 3$.

The models that we consider are of the general form of $\dot{M} =
\nu(M) M$ where $\nu(M)$ is the decay rate of a cluster with mass $M$
\citep[cf.][]{FZ}.  This form is useful, as it allows a number of
previously published evolution models to be analyzed in the same way.
Generally, these models have $\nu(M) \propto M^{-\gamma}$, or $\dot{M}
\propto M^{1 - \gamma}$.  The first such model we consider is a
standard evaporation model in which $\dot{M} \propto M^{0}$, such as
the one described by \citet{FZ}.  We also consider a simplified form
of the evaporation model derived by \citet{BM} from their analysis of
N-body simulations of globular clusters.  Briefly, their result was
that the disruption timescale for clusters scales as $T_{dis} \propto
M^{0.75}$, which we can convert into a mass derivative by noting that
$\int_{M_0}^0 \dot{M}^{-1} dM = T_{dis}$, and then solving this
equation for a mass loss rate as defined above.  This gives a final
derivative of $\dot{M} \propto M^{0.25}$.  Finally, we investigate the
numerical fit to this result that is presented in \citet{Lamers}.
They also present a disruption time, which they solve for in terms of
the cluster mass and local density.  We can convert this disruption
time, $T_{dis} \propto M^{0.62}$, into a mass loss rate $\dot{M}
\propto M^{0.38}$.

As these models only evolve single clusters, we need to create a
population of globular clusters to compare with the observed mass
function.  We used two initial mass function models that have been
used to fit the high mass globular cluster mass function before.  As
these clusters are not affected greatly by dynamical evolution, they
are assumed to trace the initial mass function closely.  The first of
these functions is a straight forward power law, $dN = M^{-\beta} 
dM$, with $\beta \sim 1.8$, as suggested by studies of young star
clusters \citep[e.g.][]{ZF1999,Whitmore2003}.  The second is a more
complex model designed to more accurately fit the observed number of
high mass clusters in other galaxies, and as such, allows that data to
influence the fitting more than it does in the case of the power law.
We chose the model for the M87 globular cluster system presented by
\citet{BS}, where $dN = M^{-3/2} \exp\left(\frac{-M}{M_C}\right) dM$
with $\log(M_C) = 6.4$, based on their fit to the bright M87 data and
our adopted mass to light ratio.

To determine how well these models fit the observed mass function, we
first generated a sample of test clusters with masses randomly drawn
from a uniform distribution in $\log(M)$.  Each test cluster was then
assigned a weight calculated from the relative number of clusters
expected from the initial mass function.  This ensures that we
consider all possible masses equally, while keeping the number of test
clusters reasonable.  For each cluster in this sample, we used the
differential equations above to let the cluster evolve to an age of 12
Gyr.  After the entire sample had been evolved, it was binned to
create the final theoretical mass function.

These theoretical mass functions were then fit to our observed data.
We allowed the functions two free parameters: $k$, the coefficient of the
mass loss differential equation, and the normalization of the initial
mass function.  The coefficient determines the rate at which the
clusters evolve, with larger constants yielding faster mass loss and
shorter lifetimes.  For the evaporation model, this coefficient is
simply the constant mass loss rate.  The normalization is independent
of the exact form of the evolution, as it merely sets the total number
of clusters present before any evolution.  We ignore stellar evolution
and gravitational shocks in our models, as these processes dominate
only the earliest parts of the lifetimes of clusters.  Because of
this, our values for $N$ assume that any losses due to these processes
have already been included.  We have also ignored the effects of
dynamical friction, which would serve to decrease the observed number
of high mass clusters.  If we restrict the data only to clusters with
mass less than $1\E{6}$, then the differences between the initial mass
functions are largely ignored, and the evolved mass function is nearly
independent of the initial one.  Without this restriction, the fits
based on the \citet{BS} initial mass function match the observed full
data set better than the simple power law mass function due to the
falloff that this mass function has at high masses.

\begin{figure}
  \figurenum{5}
  \plotone{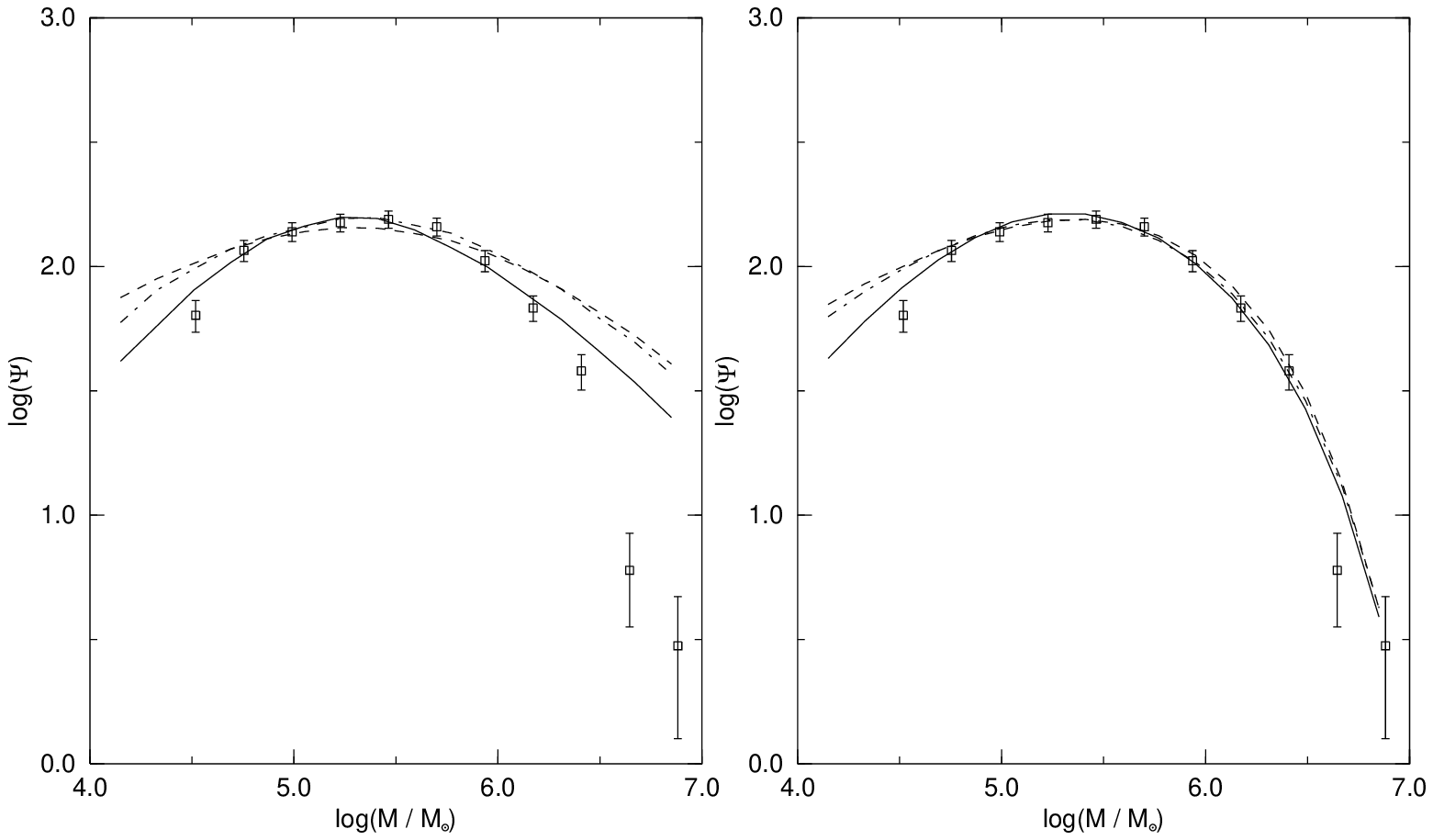}
  \caption{Comparison of our globular cluster mass function to the
    best fitting models.  The left panel assumes a powerlaw initial
    mass function, and the right panel assumes the \citet{BS} initial
    mass function.  In both plots, the solid line shows a simple
    evaporation model with $\dot{M} \propto M^{0}$, the dot dashed
    line a model with $\dot{M} \propto M^{0.25}$, and the dashed line
    a model with $\dot{M} \propto M^{0.38}$. \label{fig:GCMF_fits}}
\end{figure}

Plots of the best fitting models are given in figure
\ref{fig:GCMF_fits}, with the corresponding parameters for these fits
listed in table \ref{tab:parameters}.  The values of $k$ are given in
units of $M_\odot \textup{Gyr}^{-1}$, and the values of $N$ give the
number of initial clusters with masses between $\log(M) = 5.5$ and
$\log(M) = 7$.  As shown in figure \ref{fig:GCMF_fits}, our data are
better fit by the standard evaporation model than the other models we
consider.  However, the uncertainties for the lowest mass bins, where
the leverage to distinguish between the models is strongest, are too
large to allow us to conclusively rule out any model.

We then compare the best fitting mass loss coefficient $k$ for each of
the various dynamical models to the prediction the models give for the
mass loss rate.  For the standard evaporation model, we adopt the mass
loss rate $k = 269 \xi_e (G \bar{\rho})^{1/2} m \ln \Lambda$,
presented by \citet{FZ}, as they explicitly give this rate constant
for their solution.  Our simplification of the \citet{BM} model makes
it difficult to test that model.  We can derive the expected rate from
the \citet{Lamers} disruption time, which yields $k =
\frac{\bar{\rho}^{1/2}}{0.62 C_{env}} (10^4)^{0.62}$.  To compare
these rates with our best fit values, we have assumed a constant
density for all of our clusters based on the M87 mass profile from
\citet{Vesperini}, calculated at our median galactocentric distance
($R_g = 4.4$kpc).

For the evaporation model, the best fitting mass loss rate is
consistent with that expected from theoretical calculations.  We find
that the fit of the \citet{Lamers} model to our M87 data requires a
value of $C_{env} = 720$ in their notation, which is close to the
value of $C_{env} = 810$ they derive from the \citet{BM} disruption
time.  Somewhat lower values of $C_{env}$, which yields a faster
disruption timescale, can be accommodated by using lower mass to light
ratios to convert the cluster luminosities into masses.  However, even
including any uncertainties in our mass to light ratio, our data are
clearly inconsistent with some of the smallest values ($C_{env} \sim
300$) suggested by \citet{Lamers} based on their analysis of young
cluster systems.  This shows that mass loss rates for two-body
processes in fully relaxed old clusters can not be extrapolated from
unrelaxed young systems.

One further effect is that toward the end of a globular cluster's
lifetime, mass segregation has moved the lowest mass stars to the
outer edges of the cluster.  These low mass stars are then
preferentially stripped from the cluster, as they are less tightly
bound.  This tends to decrease the $M / L$ ratio of the cluster at the
end of its life, as these low mass stars contribute more mass to the
cluster than light.  Since the globular cluster mass function is based
on observations of the cluster light, using a constant $M / L$ ratio
for all clusters overestimates the mass of these faintest clusters.
We may see evidence for this effect, as our lowest mass bin falls
below all of the model predictions.  Any correction for such a
changing $M / L$ ratio will serve to spread our lowest mass bin to
even lower masses, which will tend to bring the result into agreement
with the predictions of the evaporation model.

\section{Discussion}
Previous studies have probed beyond the turnover of the globular
cluster luminosity function, yet none have reached the same depth as
our new luminosity function for M87.  This new depth has allowed us to
trace the effects of dynamical evolution down to clusters that are in
the last few Gyr of their lifetime, where the effects of the mass loss
are most severe.  The superior resolution offered by the data has also
allowed us to establish that the size distribution of globular
clusters is largely independent of the cluster luminosity over the
entire range we observe.

The differences in dispersion that we have observed between the
luminosity functions for M87 and the Milky Way shows that the
individual formation histories of the two galaxies may have influenced
their current shapes.  The color bimodality seems to justify the
change in dispersion due to the shift between the peaks of the blue
and red population luminosity functions.  The roughly equal number of
clusters in both populations ensures that the total luminosity
function of M87 has a wider dispersion than for the Milky Way, which
is dominated by the blue cluster population.  

Our data clearly shows that a deep luminosity function for globular
cluster systems can reliably constrain the form of the dynamical
evolution of globular clusters.  We have found that the classical
evaporation of \citet{FZ} fits our data for the central region of M87
well, as they found for the Galactic globular cluster system.  The
smallest mass clusters in our data do seem to deviate from the model,
which may be evidence that preferential mass loss from low mass stars
at the end of the cluster's life alters the mass to light ratio.  Our
error bars are not quite small enough to completely rule out other
forms for the cluster mass loss that have been proposed, although
these other models tend to fit our very deep data less well.  Our data
clearly rules out mass loss rates much more rapid than expected from
classical evaporation.

We gratefully acknowledge support for this work from grant GO-8592
from the Space Telescope Science Institute and NSF award AST-0406891
from the National Science Foundation.  We also thank E.  Vesperini for
helpful discussions, and the anonymous referee for useful comments and
suggestions.

\clearpage
\begin{deluxetable}{lcccc}
  \tabletypesize{\small}
  \tablewidth{0pc}
  \tablecaption{Number of Objects Detected\label{tab:detection}}
  \tablehead{\colhead{} & \colhead{F606W} & \colhead{F814W} & 
    \colhead{Common to both Filters} & \colhead{$0.5 < V - I < 1.7$}}
  \startdata
  Candidate Clusters& 1667 & 1918 & 1160 & 1013 \\
  Detections from Noise& 480 & 734 & 143 & 18 \\
  Detections from HDF image& 69 & 49 & 30 & 12
  \enddata
\end{deluxetable}

\begin{deluxetable}{lcccc}
  \tabletypesize{\small}
  \tablewidth{0pc}
  \tablecaption{Best Fitting Evolution Model Parameters\label{tab:parameters}}
  \tablehead{ & \multicolumn{2}{c}{Powerlaw IMF} & 
    \multicolumn{2}{c}{Burkert \& Smith IMF} \\
\colhead{Model}&
    \colhead{k}&\colhead{N}&\colhead{k}&\colhead{N}}
  \startdata
  $\dot{M} \propto M^{0}$ & 
  $1.3\E{4}$ & $1.9\E{5}$ & $1.1\E{4}$ & $1.0\E{4}$ \\
  $\dot{M} \propto M^{0.25}$ &
  $1.2\E{3}$ & $3.2\E{5}$ & $1.0\E{3}$ & $1.5\E{4}$ \\
  $\dot{M} \propto M^{0.38}$  &
  $3.3\E{2}$ & $4.1\E{5}$ & $4.0\E{2}$ & $3.2\E{4}$ 
  \enddata
\end{deluxetable}
\clearpage

\clearpage
\clearpage
\clearpage
\clearpage
\clearpage
\clearpage

\end{document}